\documentstyle[aps,pre,epsfig]{revtex}

\begin{document}

\title{Experimental evidence of chaotic advection in a
convective flow}
\author{G .Boffetta$^{1,2}$, M. Cencini$^{3,4}$, S. Espa$^{5}$
	and G. Querzoli$^{6}$}
\address{$^1$ Dipartimento di Fisica Generale, Universit\`a di Torino,
         Via P. Giuria 1, 10125 Torino}
\address{$^2$ Istituto Nazionale di Fisica della Materia,
	  Unit\`a di Torino Universit\`a, Italy}
\address{$^3$ Dipartimento di Fisica, Universit\`a ``La Sapienza'',
         P.le Aldo Moro 2, 00185 Roma}
\address{$^4$ Istituto Nazionale di Fisica della Materia,
	  Unit\`a di Roma I, Italy}
\address{$^5$ Dipartimento di Idraulica Trasporti e Strade, 
	Universit\`a ``La Sapienza'', 
	Via Eudossiana 18, 00185 Roma, Italy}
\address{$^6$ Dipartimento di Ingegneria del Territorio,
	Universit\`a di Cagliari,
	P.zza d'Armi, 90123, Cagliari, Italy}

\maketitle

\begin{abstract}
Lagrangian chaos is experimentally investigated in a convective flow
by means of Particle Tracking Velocimetry. The Finite Size
Lyapunov Exponent analysis is applied to quantify dispersion
properties at different scales.
In the range of parameters of the experiment, Lagrangian motion 
is found to be chaotic.
Moreover, the Lyapunov depends 
on the Rayleigh number as ${\cal R}a^{1/2}$.  
A simple dimensional argument for explaining the observed power
law scaling is proposed.
\end{abstract}
\vspace{0.5cm}


The investigation of transport and mixing of passive tracers is of
fundamental importance for many geophysical and engineering
applications \cite{moffat83,CFPV91}. It is now well established, and
confirmed by several numerical \cite{aref84} and experimental
\cite{SG88} evidence, that even in very simple Eulerian flow
(i.e. laminar flow) the motion of Lagrangian tracers can be very
complex due to Lagrangian Chaos \cite{CFPV91,ottino89}.  In such
a situation, diffusion may be of little relevance for transport which
is, on the contrary, strongly enhanced because of chaotic advection
\cite{aref84,ottino89}.

In this Letter, we address the problem of quantifying the dispersion
of passive tracers in a relatively simple convective flow at various
Rayleigh numbers ${\cal R}a$.  By applying the Finite Size Lyapunov
Exponent (FSLE) \cite{ABCPV96} analysis to the Lagrangian trajectories
obtained from Particle Tracking Velocimetry (PTV) technique, we are
able to estimate the dispersion properties at different scales.  We
find a clear power law dependence of the Lagrangian Lyapunov exponent
on the Rayleigh number. This dependence is explained by a dimensional
argument which excludes a role of diffusion in the dispersion process.

The experiment is performed in a rectangular tank $L=15.0\, cm$ wide,
$10.4\, cm$ deep and $H=6.0 \,cm$ height, filled with water.  Upper
and lower surfaces are kept at constant temperature while the side
wall are adiabatic.  The convection is generated by an electrical
circular heather of radius $0.4 \, cm$ placed in the mid-line of the
tank at $0.4 \, cm$ above the lower surface.  The heather works at
constant heat flux $Q$ which is controlled by a feedback on the power
supply.  By changing the heat input we control the
Rayleigh number ${\cal R}a = (g \beta Q H^3)/(\alpha \nu \kappa)$,
where $g$ is the gravitational acceleration, $\beta$ the thermal
expansion coefficient, $\alpha$ the thermal conductivity, $\nu$ the
kinematic viscosity and $\kappa$ the thermal diffusivity.  In the
parameters range explored in our experiments the flow consists of two
main counter-rotating rolls divided by an ascending thermal plume
above the heat source \cite{MQR98}. The upper end of the plume
oscillates horizontally almost periodically with a frequency which
depends on the Rayleigh number.

Lagrangian trajectories are obtained by PTV technique \cite{Q96}.
The fluid is seeded with a large number of small 
($50 \,\mu m$ in diameter), non-buoyant particles. 
The vertical plane in the middle of the tank and orthogonal to the
heat source is illuminated by a thin laser light sheet.
Single exposure images are taken by a CCD camera and 
subsequently digitalized at $8.33 \,Hz$ rate with a $752
\times 576$ pixels resolution.
Trajectories are then identified as time ordered series of particle
locations.

Each run lasts for $2700\, s$, corresponding to $22500$ frames.
Typically $900$ particles are simultaneously tracked for each frame.
In the following we analyze  the trajectories obtained in $6$
different runs with Rayleigh number in the range
$6.87 \, 10^7 < {\cal R}a < 2.17 \, 10^9$.

The analysis of Lagrangian data have been done with the Finite Size
Lyapunov Exponent tool which, introduced in the context of the
predictability problem turbulence \cite{ABCPV96}, has already been
demonstrated very efficient for the characterization of dispersion
in bounded domains \cite{ABCCV97} and in the treatment of experimental data
\cite{LAV99}.  Let us recall the basic ideas on the FSLE; more
details can be found in \cite{ABCPV96,ABCCV97}.  The idea of the
Finite Size Lyapunov Exponent is to generalize the Lyapunov exponent,
which measures the average rate of divergence of two infinitesimally
close trajectories, to finite separations.  To this aim, we fix a set
of thresholds $R_n=R_0 \rho^n$ ($n=0,\dots,N$) and we consider, at
each time $t$, new couples of trajectories ${\bf x}_1(t)$, ${\bf
x}_2(t)$ found at separation $R(t)=|{\bf x}_1(t) - {\bf x}_2(t)| <
R_0$.  We follow the evolution of the trajectories and compute the
``doubling time'' $T_{\rho}(R_n)$ it take for the separation to grow
from scale $R_n$ up to $R_{n+1}=\rho R_n$.  Of course it must be $\rho
> 1$, but we take $\rho$ not too large in order to avoid contributions
from different scales.

After performing a large number of doubling time experiments
(over the possible different couples in the run)
we average the doubling time at each scale $R$ from which
we define the Finite Size Lyapunov Exponent 
\begin{equation}
\lambda(R) = {1 \over \langle T(R) \rangle_e} \ln \rho \; ,
\label{eq:1}
\end{equation}
where $<[\dots]>_e$ indicates the average performed on the doubling
time experiments (see Refs. \cite{ABCPV96,ABCCV97} for further details).
The FSLE is a generalization of the Lyapunov exponent $\lambda$
\cite{ABCPV96}, in the sense that
\begin{equation}
\lim_{R \to 0} \lambda(R) = \lambda \; ,
\label{eq:2}
\end{equation}
physically speaking $\lambda(R) \approx \lambda$ for $R \leq l_E$, where
$l_E$ is the smallest Eulerian characteristic length.  For larger
values of $R$, $\lambda(R)$ gives information on the mechanism
governing the dispersion at  scale $R$. For example, in the case of
standard diffusion, on the basis of dimensional considerations
one expects that
\begin{equation}
\lambda(R) \simeq D/R^2\,,
\label{eq:diff}
\end{equation}
where $D$ is diffusion coefficient.

The use of the FSLE is particularly useful for studying
the dispersion properties of passive tracers in closed basins \cite{ABCCV97}, 
and, therefore, fits very well with the considered flow.
In such a situation, asymptotic regimes like diffusion (\ref{eq:diff}) might
never been reached due to the presence of boundaries. 
Denoting by $R_{max}$ the average couple separation in the asymptotic 
uniform distribution, it has been found that for a large
class of systems, for $R$ close to $R_{max}$, $\lambda(R)$ fits
the following universal behavior 
\begin{equation}
\lambda(R) \simeq {1 \over \tau_R} {R_{max}-R \over R}\,,
\label{eq:3}
\end{equation}
where $\tau_R$ has the physical meaning of the characteristic time 
of relaxation to the uniform distribution.
Equation (\ref{eq:3}) can be obtained assuming an exponential relaxation of 
tracers' concentration to the uniform distribution \cite{ABCCV97}.

We applied the FSLE analysis to the Lagrangian trajectories
experimentally obtained. 
In order to increases the statistics at large separations $R$, 
we have computed $\lambda(R)$ for different values of the smallest 
scale $R_{0}$ ($R_{0}=0.4\,cm\,,\;0.6\,cm\,,\;0.8\,cm\,$).
The threshold ratio is $\rho=1.2$ for all the analysis.
For the results presented below we use $H=6 \,cm$ as unit length
and the diffusive time $t_{\kappa}=H^2/\kappa \simeq 25000 \,s$
as unit time.

\begin{figure}[htb]
\centerline{\epsfig{figure=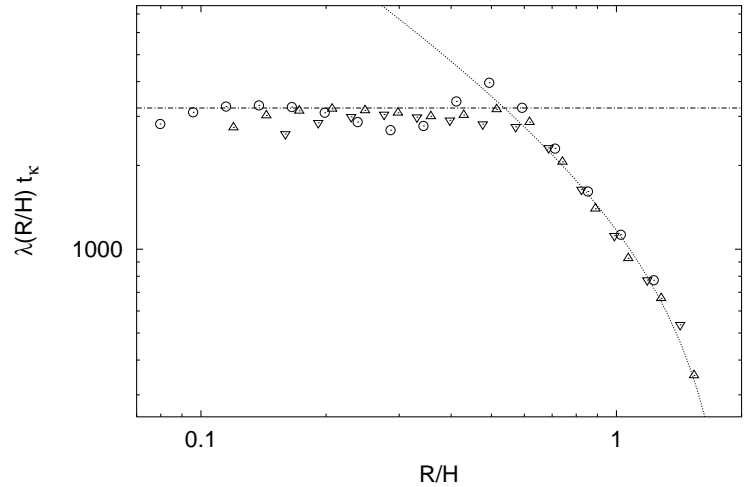,width=10cm,angle=0}}
\caption{$\lambda(R)$ versus $R$ for different initial thresholds
$R_0=0.067\,H\,\, (\circ)\, ,\;0.1\,H\,,(\bigtriangleup)\,,
\;0.13\,H \, \,(\bigtriangledown)$ at ${\cal R}a=2.39\,10^8$.
The straight line is the Lyapunov exponent
$\lambda=3100 \pm 200 \,t_{\kappa}^{-1}$ and
the curve is the saturation regime (\ref{eq:3}) with
$\tau_R=8.0 \, 10^{-4} \,t_{\kappa}$ and $R_{max}=1.9\,H$.
}
\label{fig1}
\end{figure}

In Figure \ref{fig1} we show the $\lambda(R)$ versus $R$ computed for
the run at ${\cal R}a=2.39 \, 10^8$. The first important result is the
convergence of $\lambda(R)$ to the constant value $\lambda \simeq 3100\,\,
t_{\kappa}^{-1}$ at small $R$. This corresponds to an exponential
divergence of close trajectories, i.e. a direct evidence of
Lagrangian chaos. The large value of the Lyapunov exponent (in unit of
inverse diffusive time) indicates that chaotic advection is the 
dominant mechanism for particle dispersion at small scales.

For larger separation $\lambda(R)$ drops to smaller values, indicating
a slowing down in the separation growth.  This is quantitatively well
described by the saturation regime (\ref{eq:3}). The collapse of the
curves at different $R_0$ confirms that sufficient high statistics is
reached even at these large scales. Fluctuation among different $R_0$
curves can be taken as an estimation of the error for $\lambda(R)$.
By fitting the large scale behavior of $\lambda(R)$ with (\ref{eq:3})
we obtain $R_{max} \simeq 1.9\, H$ and $\tau_{R} \simeq 8.0 \, 10^{-4}
t_{\kappa}$. Thus also for the late stage of relaxation to the uniform
distribution, diffusion seems to play a marginal role.  The
characteristic Eulerian scale in the flow $l_E$ can be estimated by
the end of the exponential regime (plateau $\lambda(R) \simeq
\lambda$) at which the non linear effects start to dominate. We find
$\l_E \simeq 0.5 \,H$, indeed not too far from the saturation value.
In this condition there is no room for the development of a diffusive
regime \cite{ABCCV97}, as Fig. 1 clearly shows.  In addition, let us
mention that by comparing $\lambda(R)$ computed at different ${\cal
R}a$, we find that both the characteristic scales $l_{E}$ and the
saturation scale $R_{max}$ are independent on the Rayleigh number.

\begin{figure}[htb]
\centerline{\epsfig{figure=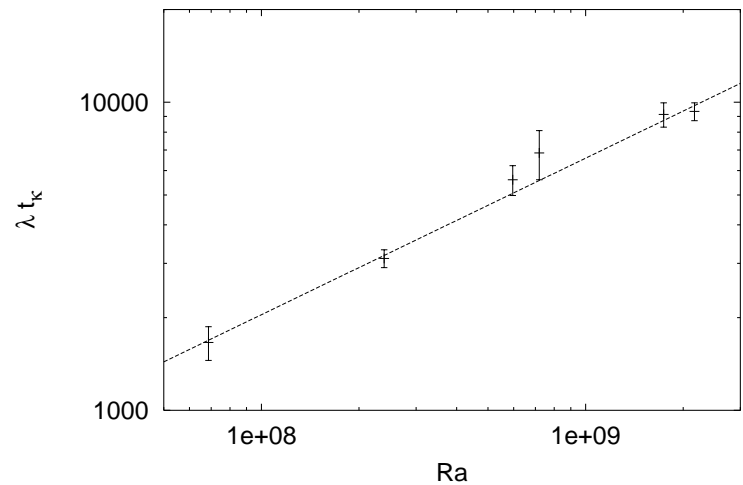,width=10cm,angle=0}}
\caption{
Lagrangian Lyapunov exponent dependence on the Rayleigh
number ${\cal R}a$.
The errors are estimated by the fluctuations
corresponding to different initial $R_{0}$.
The line is the best fit $\lambda \sim {\cal R}a^{0.51}$.
}
\label{fig2}
\end{figure}

In order to explore the dependence of the Lagrangian statistics on the
Eulerian characteristics, we have performed the FSLE analysis for
Rayleigh number varying over more than one order of magnitude.
The dependence of the Lagrangian Lyapunov exponent $\lambda$
(computed from the plateau of $\lambda(R)$ at small $R$)
on ${\cal R}a$ is shown in Figure \ref{fig2}. A clear scaling is observed, 
indicating a power law dependence
\begin{equation}
\lambda \sim {\cal R}a^{\gamma}
\label{eq:4}
\end{equation}
with $\gamma = 0.51 \pm 0.02$.
 
It is worth noting that because of the geometry of our experiment, the
flow shows qualitatively the same pattern for the whole range of
${\cal R}a$ explored. This is confirmed by the independence of $l_E$
and $R_{max}$ on ${\cal R}a$ as discussed above. In these conditions,
the scaling of $\lambda$ on ${\cal R}a$ can be supported by the following
dimensional argument. The equations of motion in the Boussinesq
approximation, made non-dimensional by $H$ and
$t_{\kappa}$, are \cite{LL87}:
\begin{equation}
{1 \over Pr} \left[
{\partial u_{\alpha} \over \partial t} +
u_{\beta} {\partial u_{\alpha} \over \partial x_{\beta}}
+ {\partial \over \partial x_{\alpha}}p  \right] =
{\partial^2 u_{\alpha} \over \partial x^2} - {\cal R}a T z_{\alpha}
\label{eq:5.1}
\end{equation}
\begin{equation}
{\partial T \over \partial t} +
u_{\beta} {\partial T \over \partial x_{\beta}} =
{\partial^2 T \over \partial x^2}
\label{eq:5.2}
\end{equation}
where $Pr=\nu / \kappa$ is the Prandtl number (which is kept constant
in our experiments).  It is now easy to verify that performing the
following rescaling:
\begin{equation}
u_{\alpha} \to \Lambda u_{\alpha}\,,\;
t \to \Lambda^{-1} t\,,\; {\cal R}a \to \Lambda^{2} {\cal R}a\,,
\label{eq:6}
\end{equation}
where $\Lambda$ is an arbitrary factor, equations
(\ref{eq:5.1}-\ref{eq:5.2}) remain unchanged, if the diffusive
terms are neglected (as it is suggested by the previous results).
Let us stress that this rescaling do not involve neither the
space (as it is suggested by above discussion) nor the Prandtl
number (because in our experiments we change $Q$, keeping
both $\nu$ and $\kappa$ constant).
The consequence of the Eulerian scaling invariance on the
Lagrangian motion, governed by
\begin{equation}
{{\rm d}{\bf x(t)} \over {\rm d}t}={\bf u}({\bf x(t)},t) \,,
\label{eq:7}
\end{equation}
is that Lagrangian trajectories are independent on the Rayleigh number.
The Lyapunov exponent, which is dimensionally
the inverse of a time, thus scales as ${\cal R}a^{1/2}$, according to
the result shown in Figure 2.

\begin{figure}[htb]
\centerline{\epsfig{figure=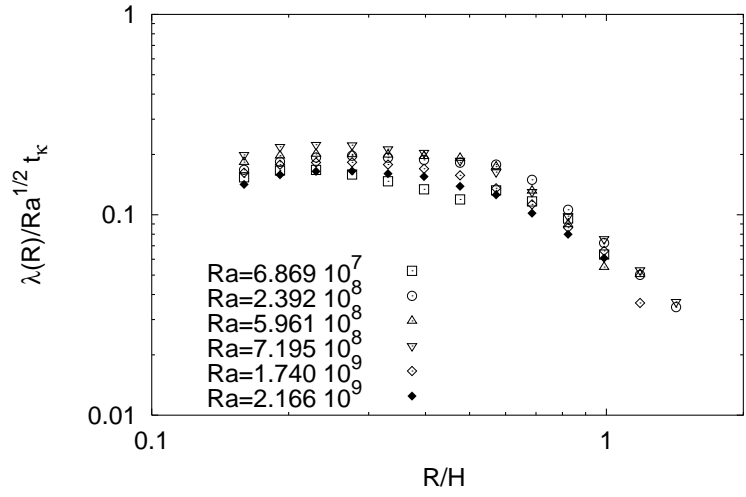,width=10cm,angle=0}}
\caption{
Data collapse of $\lambda(R)$ at different ${\cal R}a$
rescaled with ${\cal R}a^{1/2}$.
}
\label{fig3}
\end{figure}

As a consequence of the dimensional argument (\ref{eq:6}-\ref{eq:7})
the FSLE should scale with ${\cal R}a$ both in the linear and in 
the saturation ranges, i.e. $\lambda(R)/{\cal R}a^{1/2}$ is a 
$Ra$-independent function.  In order to verify the scaling argument, 
in Figure \ref{fig3} we plot the FSLE compensated with ${\cal R}a^{1/2}$
for the different runs. The collapse is fairly good, and the fluctuations
are of the order of the error reported in Figure \ref{fig2}.
As a further argument in favor of the scaling (\ref{eq:6}), let us 
observe that the transitions to the saturation regime occurs at a scale
independent on ${\cal R}a$.

As a conclusion, let us stress again that the approximate agreement
with the dimensional scaling argument and the large values of the
Lyapunov exponent in units of diffusion time give a coherent picture
of Lagrangian dispersion ruled by chaotic advection.

\vspace{0.5cm}
We thank A. Celani, A. Cenedese, S. Ciliberto, J.F. Pinton, G.P. Romano 
M. Tardia and A. Vulpiani for useful discussions.
This work is partially supported by INFM (Progetto Ricerca Avanzata TURBO) 
and by MURST.


\newpage
\section{Figure captions}

\end{document}